\documentclass[galaxies,article,accept,moreauthors,pdftex]{Definitions/mdpi}

\firstpage{1} 
\pubvolume{xx}
\issuenum{1}
\articlenumber{5}
\pubyear{2019}
\copyrightyear{2019}
\history{Received: 30 January 2019; Accepted: 11 March 2019; Published: 15 March 2019}
\usepackage{amssymb}
\usepackage{amsmath}

\usepackage{amsmath}

\DeclareMathOperator{\arctanh}{arctanh}

\setitemize{parsep=6pt,itemsep=0pt,leftmargin=*,labelsep=5.5mm}
\setenumerate{parsep=6pt,itemsep=0pt,leftmargin=*,labelsep=5.5mm}
\setlist[description]{itemsep=0mm}

\Title{What if Newton's Gravitational Constant Was~Negative?}

\Author{Ismael Ayuso %
 $^{\dagger}$, José P. Mimoso $^{*,\dagger}$ 
and Nelson J. Nunes $^{\dagger}$}

\AuthorNames{Ismael Ayuso, José P. Mimoso and Nelson J. Nunes}

\address[1] {%
Departamento de F\'{\i}sica and Instituto de Astrof\'{\i}sica e Ci\^encias do Espa\c co,
Faculdade de Ci\^encias,  Universidade de Lisboa, Edif\' icio C8, Campo Grande, 1769-016 Lisboa, Portugal;  iayuso@fc.ul.pt (I.A.); njnunes@fc.ul.pt~(N.J.N.)
}

\corres{\hangafter=1 \hangindent=1.0em \hspace{-0.6em}Correspondence: jpmimoso@fc.ul.pt
}

\firstnote{\hangafter=1 \hangindent=1.0em \hspace{-0.6em}These authors contributed equally to this work.}

\abstract{In this work, we seek a cosmological mechanism that may define the sign of the effective gravitational coupling constant, {\em G}. To this end, we consider general scalar-tensor gravity theories as they provide the field theory natural framework for the variation of the gravitational coupling. We~find that models with a quadratic potential naturally stabilize the value of {\em G} into the positive branch of the evolution and further, that de Sitter inflation and a relaxation to General Relativity is easily attained.}


\begin{document}


\section{Introduction}
In Newton's law of gravitation, the gravitational constant, $G$, is assumed to be positive. 
This is a question of choice, and apparently it was P. S. Laplace who introduced the constant  for the first time in his Traité de Mécanique C\'eleste, in 1799 \cite{Laplace 1799} as
\begin{equation}
F = - k^2\,\frac{m_1\,m_2}{r^2} \; .
\end{equation}

Originally, Newton had  put forward both the proportionality  of the gravitational centripetal force (in his words)  to the quantity of matter of the two bodies in interaction, as well as the inverse proportionality to the square of their separation \cite{Newton:1687eqk}. However,  he did not explicitly introduce $G$~\cite{Will:1986nk}\footnote{Quoting Clifford Will~\cite{Will:1986nk}, "It is interesting to notice that the term ``{\em gravitational constant}'' never occurs in the {\em Principiae}. In fact it  seems that the universal constant of proportionality that we now call $G$ does not make an appearance until well in the eighteenth century in Laplace's ``M\'ecanique C\'eleste''.}, presumably due to the lack of an internationally accepted  system of units.  Of course, this is required since, after all, $G$ adjusts the dimensions of both sides of the defining equation for the strength of the gravitational interaction and the sign of this denotes the attractive or repulsive character of the force. Odd enough, it was in 1798, one year before the publication of Laplace's treaty, that H. Cavendish measured $G$ with a torsion balance, but just as a necessary step, of secondary importance, to weigh the density of the Earth \cite{Eotvos-Wash I}. This measure was made with  the remarkable accuracy of $1\%$. 

The subsequent success of the gravitational law in tackling the motion of the celestial bodies of the solar system  is well known, and, at the beginning of the 20th century, the only major problem was 
the anomaly  in the  precession of Mercury's perihelium, a mismatch first revealed by  Le Verrier  in 1855.

It was Einstein's General theory of Relativity which not only solved this puzzle with flying colours, but also revolutionized our understanding of gravitation. One of the pillars of the theory is that we should recover Newtonian gravity when considering weak fields and bodies moving with low speeds when compared to the speed of light. Thus, $G$ whose role  in the theory is to  couple the geometry to the matter content of the Universe, is taken to be positive and is a constant under this framework. 


In 1938, Dirac made an astounding proposal, dubbed the Large Number Hypothesis, according to which any {\em dimensionless} ratio between two fundamental quantities of nature 
should be of the order unity (for a more detailed  account of the motivations  see \cite{Barrow:1988yia}). This led him to put forward that if $G$ were to evolve with the Hubble rate of expansion of the Universe this would account for the present disparity of about 40 order of magnitude between gravitational and electromagnetic forces at the atomic level. 

This was the first time the variation of some fundamental constant was explicitly and seriously envisaged. Dirac's proposal was  given a field theoretical realization, first by P. Jordan  within a Kaluza-Klein type approach (thus involving extra-dimensions), and then, in 1961, by Brans-Dicke theory~\cite{Will:1986nk,Brans:1961sx} motivated by Mach's principle.  In both cases a dynamical scalar field couples to the spacetime curvature and thus plays itself a gravitational role. 
In the suite, a plethora of extended gravity theories that affect the coupling between the space-time  geometry and the matter sector  also prescribe the variation of this fundamental ``constant'' $G$ \cite{Clifton:2011jh,Avelino:2016lpj}.   

In  principle, within this framework, it becomes possible for $G$ to change sign, trading, attracting into repulsive gravity, and conversely.  This might happen either during the cosmological time evolution, or even conceivably  it might happen at spatially separated regions of space-time. The~concern about the sign of the gravitational constant has been envisaged as a constraint to be respected by the spectrum of modified gravity theories, but the focus has never been directed to devise a mechanism to assure its positiveness. For instance, Barrow \cite{Barrow:1992ay} proposed that the formation of primordial black holes during the early stages of the Universe might retain ``memory'' of the value of the gravitational constant at the time of their formation, and hence exhibit diverse values of the latter depending on the instant of their formation, around $t_{Prim}\sim 10^{-25}$ s.
 
In the present work we investigate the, somewhat heretical, possibility  that the effective gravitational coupling might be negative within the general class of scalar-tensor (ST) gravity theories. We analyze a  cosmological mechanism that determines the positiveness of the sign of G, even though it may exhibit transient periods in the negative region. We show that this cosmological device relies on the role of a cosmological potential, which reproduces a positive cosmological constant  in the so-called Einstein frame. From this latter viewpoint it can be understood as another role of paramount importance of this remarkable constant. 
In Refs. \cite{Roxburgh:1980,Roxburgh:1980b} I. Roxburgh analysed the issues of the sign and magnitude of the gravitational constant, based on Einstein's correspondence principle which demands that Newtonian gravity be recovered in the weak field limit of the theory. His analysis is done in the framework of GR and is somewhat motivated by Mach's principle,  leading him to conclude that $G$ must be positive. Other studies which carry some relation to the present work are \cite{1982PhRvD..26.2664K,1978ApJ...219....5B, Mimoso:2003iha,Mimoso:1998dn,Mimoso:1999ai,Mimoso 2011spr,Jarv:2010zc,Jarv:2010xm}.

In this work we shall start by briefly looking at  the implications of having  $G<0$ in cosmology, namely showing that inflation arises for a considerably large set of parameters, and that we obtain bouncing solutions that avoid the initial singularity when a cosmological constant is also considered. Then we analyze the cosmological behaviour of scalar-tensor theories to show how a subset of the solutions exhibit negative $G$, and how  a cosmological potential  provides us with a mechanism that favours positive $G$ and eventually stabilizes its sign. In essence we will show that the presence of a cosmological constant in the Einstein frame  provides such a mechanism for an extended set of varying $G$ theories, which represents  a relevant feature for the existence of a non-vanishing cosmological constant in the Einstein frame (and of a corresponding cosmological potential in the Jordan frame).

\section{Negative \emph{G} in GR}

It must be said that if we envisage the trading of a positive $G$ into a negative one within Einstein's General Relativity, we will be mutating its attractive nature into  a repulsive one, and this avoids the need to rely on exotic matter, violating the strong energy condition, to produce inflationary stages. This  is therefore  an alternative ad-hoc device, akin to the Albrecht and Magueijo's varying speed  of light to avoid the perplexing complications of the inflationary scenarios \cite{Albrecht:1998ir}. The~down side of this way of producing repulsive gravity, is that once assumed, it is for ever.  There would be no way of exiting inflation with canonical matter sources.  The scalar-tensor scenario that we consider afterwards avoids the latter problem, and present us with a natural, and theoretically consistent framework for exploring the possible negativeness of $G$.

\subsection{Friedmann Models with a Single Fluid}
Consider the usual FLRW universes of the standard cosmological model, and take $G=-|G|$ in Einstein's GR.  We then have the following field equations
\begin{eqnarray}
\frac{\dot a^2}{a^2}+\frac{k}{a^2}&=& -\frac{8\pi |G|}{3}\,\rho \label{eq_Fried}\\
\frac{\ddot a}{a} &=&  \frac{8\pi |G|}{6}\,(\rho+3p)\; , \label{eq_Rauch}
\end{eqnarray}
where dots denote derivatives with respect to the time, $a$ is the scale factor of the Universe and $\rho$ and $p$ are the energy density and pressure, respectively. The~signs on the right hand side are the opposite with respect to the usual ones. 
However, the Bianchi contracted identities are immune to this change of sign and the energy conservation equation is preserved
\begin{equation}
\dot{\rho} = - 3H\,(\rho+p) \; . \label{eq_ener_cons}
\end{equation}

Thus, when the matter content satisfies the weak and strong energy conditions, $\rho>0$, $\rho+p\ge 0$, and~$\rho+3p\ge 0$, we see from the Raychaudhuri Equation,~(\ref{eq_Rauch}), that the expansion is accelerated, $\ddot a \ge 0$.  Yet,~this inflationary behaviour is constrained 
by the Friedmann Equation (\ref{eq_Fried}). It can be easily verified that the single fluid solutions are forbidden when $k=0,+1$, and are restricted to $\rho\le 3/8\pi|G|a^2$ when~$k=-1$. 

Further, notice that the transformation $|G| \to -|G|$ which is performed in the Einstein field equations of the FLRW models, produces a system which mimics phantom matter provided the equation of state relating the pressure and the energy  density of matter is such that
$p(\rho)\to -p(-\rho)$ when $\rho \to -\rho$, preserving the field equations  
(we remark that this happens to be the case for the barotropic equations $p=(\gamma-1)\rho$ which are usually considered; in addition the cosmography framework, as exposed in  \cite{Dunsby:2015ers, Aviles:2012ay}, also absorbs this transformation and is left unchanged).

\subsection{Model with a Cosmological Constant}

Consider a cosmological constant in addition to the perfect fluid 
\begin{equation}
T_{ab} = -\Lambda \,g_{ab}\; .
\end{equation}
for a metric with the signature $-+++$. The~field equations now read
\begin{eqnarray}
\frac{\dot a^2}{a^2}+\frac{k}{a^2}&=& \lambda-\frac{8\pi |G|}{3}\,\rho \label{eq_Fried-Lambda}\\
\frac{\ddot a}{a} &=&  \lambda + \frac{8\pi |G|}{6}\,(\rho+3p)\; , \label{eq_Rauch-Lambda}
\end{eqnarray}
where $\lambda=\Lambda/3$.
Recasting the latter equations in conformal time $\eta$ defined by ${\rm d}\eta= {\rm d}t/a(t)$ we get
\begin{eqnarray}
(a')^2+{k}{a^2}&=& \lambda\,a^4-\frac{8\pi |G|}{3}\,\rho\,a^4 \label{eq_Fried-Lambda_CT}\\
\frac{a''}{a} &=&  2\lambda\, a^2-{k}- \frac{8\pi |G|}{6}\,(\rho-3p) \,a^2\; , \label{eq_Rauch-Lambda_CT}
\end{eqnarray}
where $'$ denotes derivative with respect to the conformal time. Assuming that the matter content is a perfect fluid with equation of state (EOS) $p=(\gamma-1)\,\rho$ where $\gamma$ is a constant that takes values in the range $0<\gamma\le 2$, we derive
the exact solutions from Equation (\ref{eq_Fried-Lambda_CT})
\begin{equation}
\int\,\frac{{\rm d}a}{\left(\lambda\,a^4-(8\pi |G|/3)\,\rho_0\,a^{4-3\gamma}-k\,a^2\right)^{1/2}} =\pm (\eta-\eta_0)
\end{equation}
which  yields Jacobi elliptic functions. Naturally, in the latter equation $\eta_0$ is an arbitrary integration constant that sets the origin of time. There are four  cases  that are of special interest: (i) Radiation, i.e., $\gamma =4/3$, (ii) Dust, i.e., $\gamma =1$, (iii) Stiff matter, i.e., $\gamma =2$, and (iv) The coasting model $\gamma = 2/3$.  

A case which is of great interest is the case where we have a combination of pressureless matter and radiation together with a cosmological constant, since these are the 3 major components that best fit the expansion history of the universe ($\Lambda$CDM model)~\cite{Aghanim:2018eyx}. 

The corresponding dynamical system, from (\ref{eq_Rauch-Lambda_CT}) reads
\begin{eqnarray}
a' &=& b \label{ds_GR_a}  \\ 
b' &=& - k\, a-\frac{8\pi |G|}{6}\rho_d^0 +2 \lambda \,a^3 
= -k a-\frac{\Omega_m^0 H_0^2}{2}+2\Omega_\Lambda^0 H_0^2 a^3, \label{accelerationGR}
\end{eqnarray}
where $\rho_d^0$ is the current density of dust. In addition, the Friedmann constraint equation becomes:
\begin{equation}
k a^2 + b^2 = -\frac{8\pi |G|\rho^0_d}{3}\,a -\frac{8\pi |G|\rho^0_r}{3} +\lambda\,a^4\; .
\end{equation}

\begin{figure}[H]
\centering
\includegraphics[height=.25\textheight=.2,angle=0]{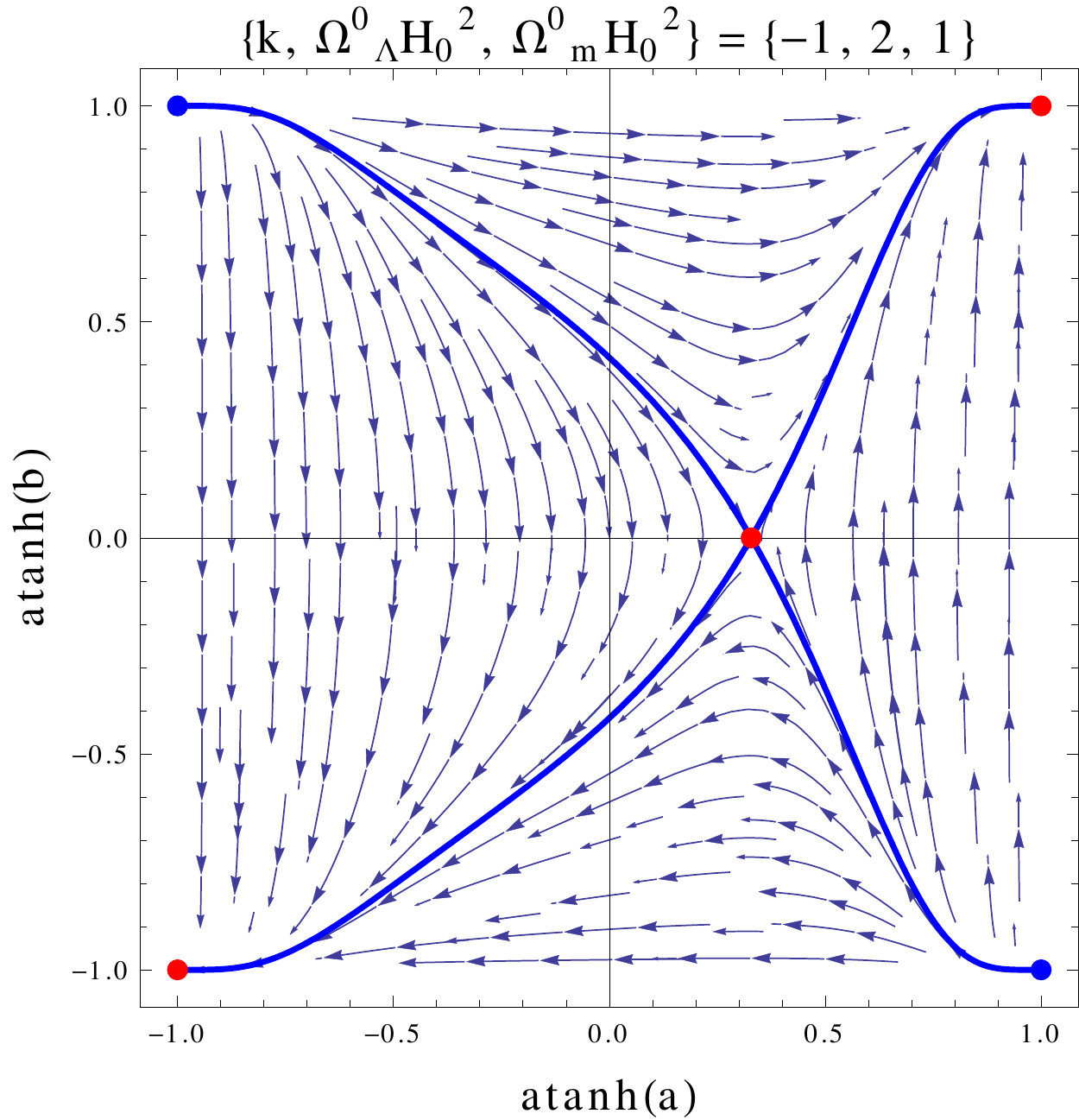}\;\;\;\;\;\;\;\;
\includegraphics[height=.25\textheight=.2,angle=0]{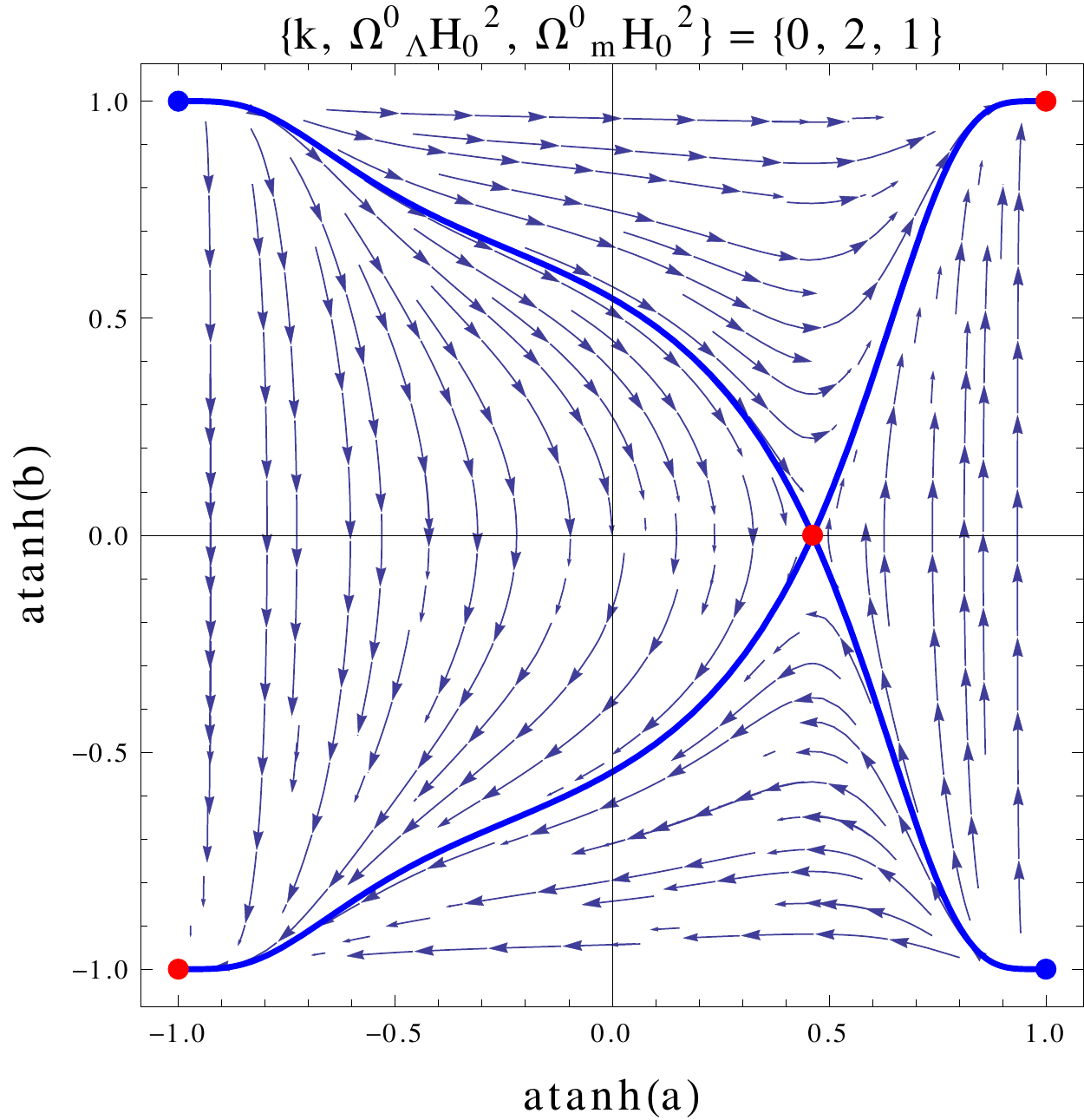}\\
\hspace{+20pt}(\textbf{a})\hspace{+165pt}(\textbf{b})\\\vspace{5pt}
\includegraphics[height=.25\textheight=.2,angle=0]{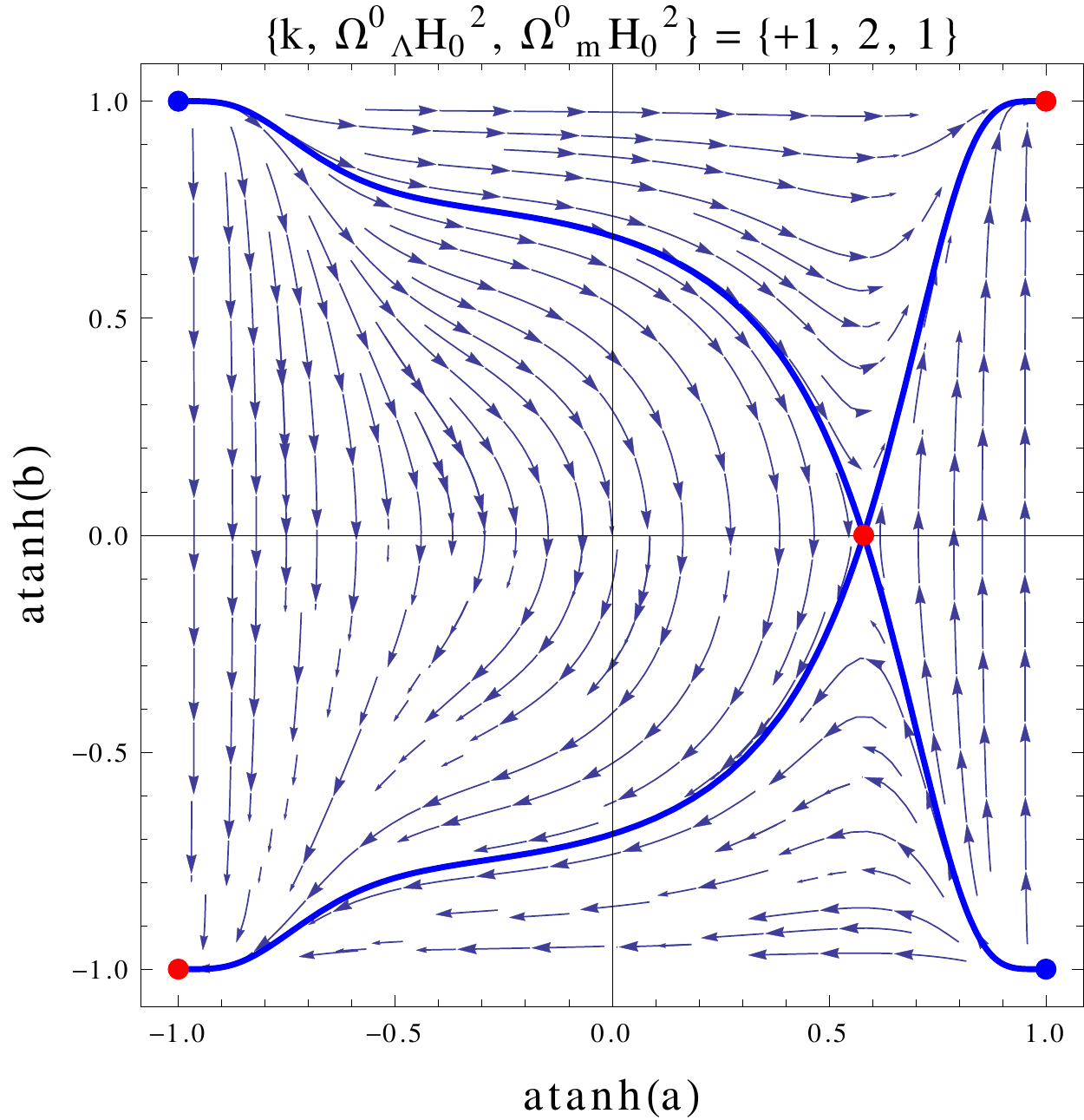}\;\;\;\;\;\;\;\;
\includegraphics[height=.25\textheight=.2,angle=0]{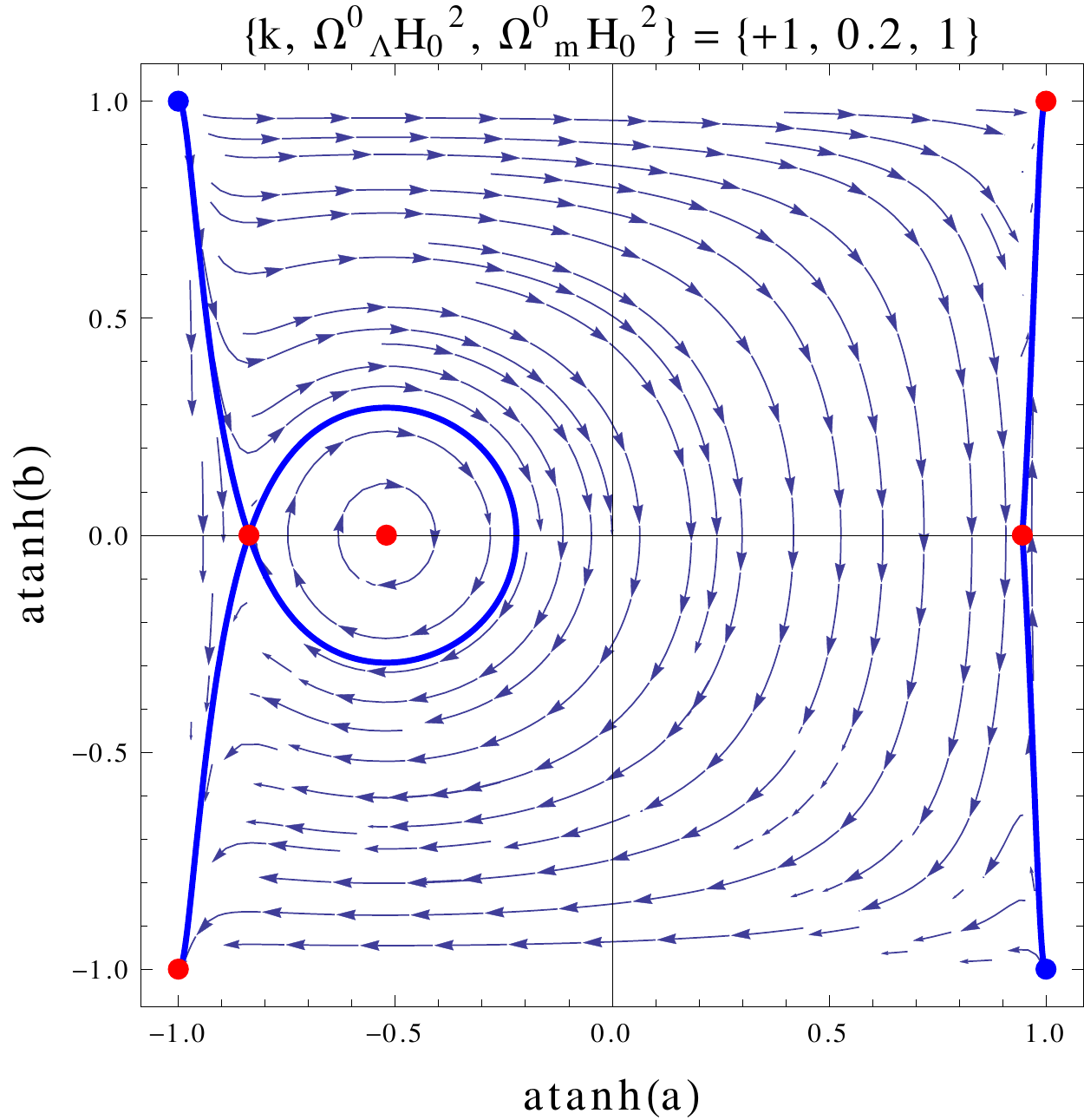}\\
\hspace{+20pt}(\textbf{c})\hspace{+165pt}(\textbf{d})
\caption{Phase-diagrams displaying the behaviour of the vector field $(a,a')$ for GR with a negative $G$. From left to right and from top to down: (\textbf{a}) represents the $k=-1$ models,  (\textbf{b}) the $k=0$ models, and~(\textbf{c},\textbf{d}) correspond both to the $k=+1$ models. 
}
\label{figureGR}
\end{figure}

 In Figure \ref{figureGR}, we represent the phase diagrams depicting the qualitative behaviour of these negative $G$ models for some choices of matter content (For a recent review of the methods of dynamical systems in cosmology see \cite{Bahamonde:2017ize}). Analyzing  the existence and nature of the fixed points, we classify the possible dynamical behaviours. Please note that we have compactified the phase diagrams using the transformation  $x=\arctanh a$ and $y=\arctanh b$, so that the boundary lines $x=\pm 1$ and $y=\pm 1$, respectively, correspond to $a\to \pm \infty$ and $b\to \pm \infty$. This allows us to devise the asymptotic solutions at infinity.
 
 The number and position of the fixed points in the finite region of the phase plane $(a,b)$ is defined by the roots of Equation (\ref{accelerationGR}) when $b=0$. Therefore, there will be at most three fixed points on the $a$ axis (plus the fixed points at infinity which will not be on the $a$ axis).  In Figure \ref{figureGR} we display the qualitative behaviour of the model for the three spatial curvatures and use reasonable values for the parameters in  Equation (\ref{accelerationGR}),
which take into consideration the $\Lambda$CDM model\footnote{One must though be wary that in the phase-diagrams of Figure \ref{figureGR} the half-plane corresponding to negative values of $a$ is not physical, as it corresponds to $a<0$. Yet its representation is useful, because it illustrates the complete behavior of the mathematical dynamical system underlying the physical scenario, regardless of the physical consistency of some of its parts. Moreover, in the present case it also allows comparison with the phase-diagrams of the scalar-tensor models.}. We adopt 
$\Omega_\Lambda^0 H_0^2/\Omega_m^0 H_0^2 \approx 2$, except in Figure  \ref{figureGR}d.

In  Figure  \ref{figureGR}a,b $k = -1$ and $k = 0$ models are represented, and Figure \ref{figureGR}c,d correspond both to the $k=+1$ models. The~lower half-planes of the various diagrams are mirror reflections from the upper ones under the $t\to -t$, $(a,b)\to (a, -b)$ symmetry. In  Figure  \ref{figureGR}a there are three fixed points, one  at $\{a,b\}=\{a_\ast,0\}$, where $a_\ast$ is a finite root of the right-hand side of Equation (\ref{accelerationGR}) (which for our choice of parameters corresponds to $\{a,b\}=\{0.341164,0\}$), 
and two at infinity. The~former fixed point is a saddle point denoting an unstable static solution ; the latter points at $b=\pm \infty$ are contracting and  expanding de Sitter solutions related by the time reversal symmetry. Although the model is open, there are re-collapsing solutions at the left of the finite fixed point, and solutions collapsing from infinity and bouncing back to it. In  Figure  \ref{figureGR}b there also are three fixed points, one  at $\{a,b\} = \{0.5,0\}$,  and the other two at $b = \pm \infty$. The~qualitative behaviour  is alike the one found for $k=-1$ case. The~Figure \ref{figureGR}c,d phase diagrams display the two possible behaviours of the $k=+1$ which translate the existence of a bifurcation associated with two different subsets in what regards the  balance of parameters (the~bifurcation occurs for $8\pi|G|\rho_d^0 =4/\sqrt{6\lambda}$). In~the left one, Figure \ref{figureGR}c, once again there are three fixed points at $\{a,b\}=\{0.662359,0\}$ plus the two  fixed points at infinity. From a qualitative viewpoint we find the same behaviour as in the previous open models. However, in Figure \ref{figureGR}d  there are three fixed points on the $x$-axis, two saddle points and a center in between, the latter of which corresponds to a basin of oscillatory behaviour. Yet~being located in the left-half plane, i.e., $a<0$, its impact on the physical  right-hand  side is not qualitatively noticeable, apart from reducing the proportion of solutions which evolve towards the deS points at $b=\pm\infty$.

Analyzing the fixed points of the compactified phase-diagrams, we find that for $k=-1$ and $k = 0$, there are only three fixed points, while for $k = +1$ there could be either  three or five fixed points, counting the two critical points at infinity. In the former cases,
when there are only three fixed points, the qualitative behaviour is the same independently of the value of $k$. Different spatial curvature indexes  only distinguish through a horizontal shift of the location of the critical point on the horizontal axis. These fixed points are saddle  points which correspond to unstable static solutions. In  all $k$ cases, there are solutions where $a$ expands to infinity. This fact was foreseeable, because of the repulsive character of gravity when $G$ is changed to  $-|G|$.  They correspond to  asymptotic de Sitter solutions (deS), both in the future and in the past, upon time reversal. They reflect the eventual domination of the cosmological $\lambda$-term, overcoming the impact of the sign of $G$, yet the swapping of the sign of $G$ enhances this domination.  Obviously, according to this behaviour, the current accelerated expansion of the Universe is not a problem, but we have a gravity which is inconsistent at small scales with the weak field  limit, and thus would be at odds with the Solar System behaviour~\cite{Roxburgh:1980,Roxburgh:1980b}. \mbox{However, the analysis pursued} in this section is merely a previous step to assess the cosmological impact of a negative gravity. In the following section we shall consider the issue within the more appropriate framework of modified metric gravity theories which assume the variation of $G$. 

\section{Scalar-Tensor Gravity Theories}

We now consider general scalar-tensor gravity theories given by the action
\begin{equation}
S=\int  \sqrt{-g}\;{\rm d}^{4}x\left[\phi\, R - \frac{\omega(\phi)}{\phi}\,\phi_{,\mu} \phi^{,\mu} - 2  U(\phi)\right]+S_{m}
,
\label{action2}
\end{equation}
where a  potential term $U(\phi)\,$ of cosmological nature is considered.
(We shall also use $U(\phi)=\phi\,\lambda(\phi)$). The~archetypal feature of this class of theories is the fact the Newton's gravitational coupling is $G=1/\phi$, and generically varies. The~scalar field $\phi$ may be seen as the gravitational permittivity of the space-time~\cite{Cembranos:2018lcs}.

The field equations are
\begin{eqnarray}
R_{\alpha \beta}-\frac{1}{2}\,g_{\alpha \beta}\, R+\lambda(\phi)\,g_{\alpha \beta}&=&\frac{\omega(\phi)}{\phi^2}
\left[\phi_{;\alpha}\phi_{;\beta} -\frac{1}{2}\, g_{\alpha \beta}\phi_{;\gamma}\phi^{;\gamma}\right]
+\frac{1}{\phi}\,\left[\phi_{;\alpha \beta}-g_{\alpha \beta}{\phi_{;\gamma}}^{;\gamma}\right]+
8\pi \,\frac{T_{\alpha \beta}}{\phi}
\label{eSTTFE1}  \\
& & \nonumber\\
\square{\phi}-\frac{2\phi^2 \lambda'(\phi)-2\phi \lambda(\phi)} {2\omega(\phi)+3}
&=&\frac{1}{2\omega(\phi)+3}\left[ 8\pi \, T-\omega'(\phi) \phi_{;\gamma}\phi^{;\gamma}
\right] \label{eSTTFE2}
\end{eqnarray}
where $T\equiv {T^c}_c\,$ is the trace of the energy-momentum tensor, 
${T_\alpha}^\beta$.
When applied to the FLRW models we obtain
\begin{equation}
3\left(\frac{\dot{a}}{a}\right)^2 +3 \frac{\dot{a}}{ a}
\frac{\dot{\phi}}{ \phi} + 
3 \frac{k}{ a^2} = \lambda(\phi)+\frac{\omega(\phi)}{ 2} 
\frac{\dot{\phi}^2}{ \phi^2}
+ 8\pi  \frac{\rho}{ \phi} ,\label{FRW-Fried_eq}
\end{equation}
\vspace{5pt}
\begin{equation}
2\frac{\rm d}{{\rm d}t}\left(\frac{\dot{a}}{ a}\right) +
3\left(\frac{\dot{a}}{ a}\right)^2 +2 \frac{\dot{a}}{ a}
\frac{\dot{\phi}}{ \phi} + 
\frac{k}{ a^2} = \lambda(\phi)-\frac{\omega(\phi)}{ 2} 
\frac{\dot{\phi}^2}{ \phi^2}
- 8\pi \frac{p}{ \phi} - \frac{\ddot{\phi}}{\phi}, \label{FRW-Ray_eq}
\end{equation}
\vspace{5pt}
\begin{equation}
\ddot{\phi}+3\frac{\dot{a}}{ a}\dot{\phi}
 + \frac{{2\phi^2\lambda'(\phi)-2\phi\lambda(\phi)}}{ {2\omega(\phi)+3}} =
 -  \frac{1}{2\omega(\phi)+3}\left[ 8\pi  (3p-\rho)+
\omega'(\phi) \dot{\phi}^2 \right] \quad.
\label{FRW-sf_eq}
\end{equation}
%

Please note that  the cosmological potential $U(\phi)=\phi\,\lambda(\phi)$ effectively reduces to a cosmological constant when $\lambda(\phi)=\lambda_0=$ constant in this frame .



We introduce the redefined variables
\begin{equation}
X=\frac{\phi}{\phi_0}\, a^2 \; , \qquad Y' = \sqrt{\frac{2\omega{(\phi)}+3}{3}}\,\frac{\phi'}{\phi}\; \quad ,
\end{equation}
and use conformal time ${\rm d}\eta= {\rm d}t/a={\rm d}\tilde t/\sqrt{X}$.
Observe that in the definition of $X$, $\phi_0$ is the value of $\phi$ at some initial condition which we shall normalize $\phi_0=1$ without loss of generality. More importantly, observe that $X<0$ when $\phi<0$, i.e., when $G<0$. This is in fact the  crucial detail which allows us to extend the study of the dynamics into the region where $\phi=1/G$ is negative.

The FLRW equations are then recast as
\begin{eqnarray}
(X')^2 + 4 \, k \,X^2 - (Y'\, X)^2 &=& 
4 M \;X \, \left(\frac{X}{\phi}\right)^{\frac{4 -3\gamma}{2}}+ \frac{4}{3} \,\left(\frac{\lambda(\phi)}{\phi}\right)\, X^3,
\label{eLPamV}
\end{eqnarray} 

The  scalar-field equation
\begin{eqnarray}
\left[Y'\,X\right]'= M(4-3\gamma)\;\sqrt{\frac{3}{2\omega+3}}\;
\left(\frac{X}{\phi}\right)^{\frac{4 -3\gamma}{2}} - \frac{2\, X^2}{\sqrt{2\omega(\phi)+3}}\;\left(\frac{{\rm d}\lambda}{{\rm d}\phi}-\frac{\lambda(\phi)}{\phi}\right),
\label{eLPbmV}
\end{eqnarray}

The  generalized Raychaudhuri equation
\begin{equation}
X'' + 4\,k \; X= 3\,M(2-\gamma)\;
\left(\frac{X}{\phi}\right)^{\frac{4 -3\gamma}{2}}
+2 X^2 \;\left(\frac{\lambda(\phi)}{\phi}\right),
\label{eLPcmV}
\end{equation}
where $M\,$ is a constant defined by $M\equiv 8\pi  \rho_0/3\,$.

When the potential $U(\phi) = \lambda_0 \,\phi^2$, the  latter equations reduce to
\begin{equation}
(X')^2 + 4 \, k \,X^2 - (Y'\, X)^2 = 
4 M \;X \, \left(\frac{X}{\phi}\right)^{\frac{4 -3\gamma}{2}}
+ \frac{4}{3} \,\lambda_0\, X^3,
\label{eLPamV_Lambda}
\end{equation} 
\begin{equation}
\left[Y'\,X\right]'=M(4-3\gamma)\;\sqrt{\frac{3}{2\omega+3}}\;
\left(\frac{X}{\phi}\right)^{\frac{4 -3\gamma}{2}}
\label{eLPbmV_Lambda},
\end{equation}
%
\begin{equation}
X'' + 4\,k \; X= 3\,M(2-\gamma)\;
\left(\frac{X}{\phi}\right)^{\frac{4 -3\gamma}{2}}
+2 \lambda_0\,X^2 .
\label{eLPcmV_Lambda}
\end{equation}

These equations can be exactly integrated  for the cases where the variables decouple, which are vacuum, radiation and stiff matter~\cite{GeneralSTsols,GeneralSTsols2,GeneralSTsols3}. Indeed, from Equation (\ref{eLPbmV_Lambda}) we see that in these cases
\begin{equation}
    Y'=\frac{f_0}{X(\eta)}\; , \label{Y_integrable}
\end{equation} 
where is an arbitrary integration constant which fixes the initial value of $Y$. However, for our present purposes,  we only need to assess the qualitative behaviour of the dynamical  system~\cite{DynSis_GenSTT3,Jarv:2010xm,Charters:2001hi,DynSis_BD1,Carloni:2007eu}.  

The crucial point in our analysis of the sign of the gravitational coupling $\phi$ is that instead of choosing either the original so-called Jordan frame or the conformally transformed Einstein frame arising from rescaling the metric with a factor $\phi/\phi_0$, where $\phi_0$ is an initial value of $\phi$, say, at present, we consider the variables $(X,Y)$, where $X=\phi a^2/\phi_0$ reflects the actual sign of $\phi$, as $a^2\geq 0$. 
 
An inspection of Equations~(\ref{eLPamV_Lambda})--(\ref{eLPcmV_Lambda}) shows that when $X\to 0$, Equation (\ref{eLPamV_Lambda}) is dominated by the scalar field term $(Y'X)^2$, which is constant for radiation, so that we expect the phase space trajectories to cross the $X=0$ axes from right to left when $X'<0$, and from left to right when $X'>0$. Interestingly, Equation~(\ref{eLPcmV_Lambda}) shows that when $X\to 0$, the dominant term is $3M(2-\gamma)\lim_{X\to 0}(X/\phi)^{(4-3\gamma)/2}$, so that it is actually the matter term the responsible for the turning around of the trajectories towards the positive side of $X$, and hence of $\phi>0$. Finally, the presence of a quadratic cosmological potential eventually dominates for large values of $X$ and consequently stabilizes the sign of $\phi$. In the following subsections we perform a qualitative analysis confirming this behaviour.

Studies of scalar-tensor theories have been performed by one of the present authors using these techniques \cite{Mimoso:1998dn, Mimoso:2003iha}, and similar and complementary analysis can be found in the literature that followed~\cite{DynSis_BD1,Jarv:2010zc,Jarv:2010xm,Carloni:2007eu,Hrycyna:2013yia,Garcia-Salcedo:2015naa}. Some of these works focus their analysis to the case of~Brans-Dicke theory \cite{DynSis_BD1,Hrycyna:2013yia,Garcia-Salcedo:2015naa}, while others investigations consider more general scalar-tensor theories. In most of the cases, the~qualitative studies rely on  choices of variables that make it difficult or even impossible
to discuss the sector of the phase space where $\phi$ is negative, e.g., \cite{Hrycyna:2013yia,Garcia-Salcedo:2015naa} (for a more detailed discussion of the use of the qualitative analysis of dynamical systems applied to scalar-tensor theories see \cite{Bahamonde:2017ize,DynSis_GenSTT3} and~references therein).

\subsection{Models without a Cosmological Potential}

We begin considering the case where the Brans-Dicke (BD) like scalar field $\phi$ is massless,\mbox{ i.e., the cosmological potential is absent.} This will be contrasted with the case where there is a quadratic potential. By the same token it will enable us to assess whether there is any effect due to a variation of the coupling $\omega(\phi)$ with regard to the issue of determining the sign of $\phi$.

For this case, the previous Equation (\ref{eLPcmV_Lambda}) can be written for vacuum ($M = 0$) and a stiff fluid ($\gamma = 2$) as:
\begin{eqnarray}
X'&=&W\nonumber\\
W'&=&-4 k X
\end{eqnarray}
and for radiation ($\gamma = 4/3$) as:
\begin{eqnarray}
X'&=&W\nonumber\\
W'&=&2M-4 k X\;.
\end{eqnarray}

\begin{figure}[H]
\centering
\includegraphics[height=.20\textheight=.2,angle=0]{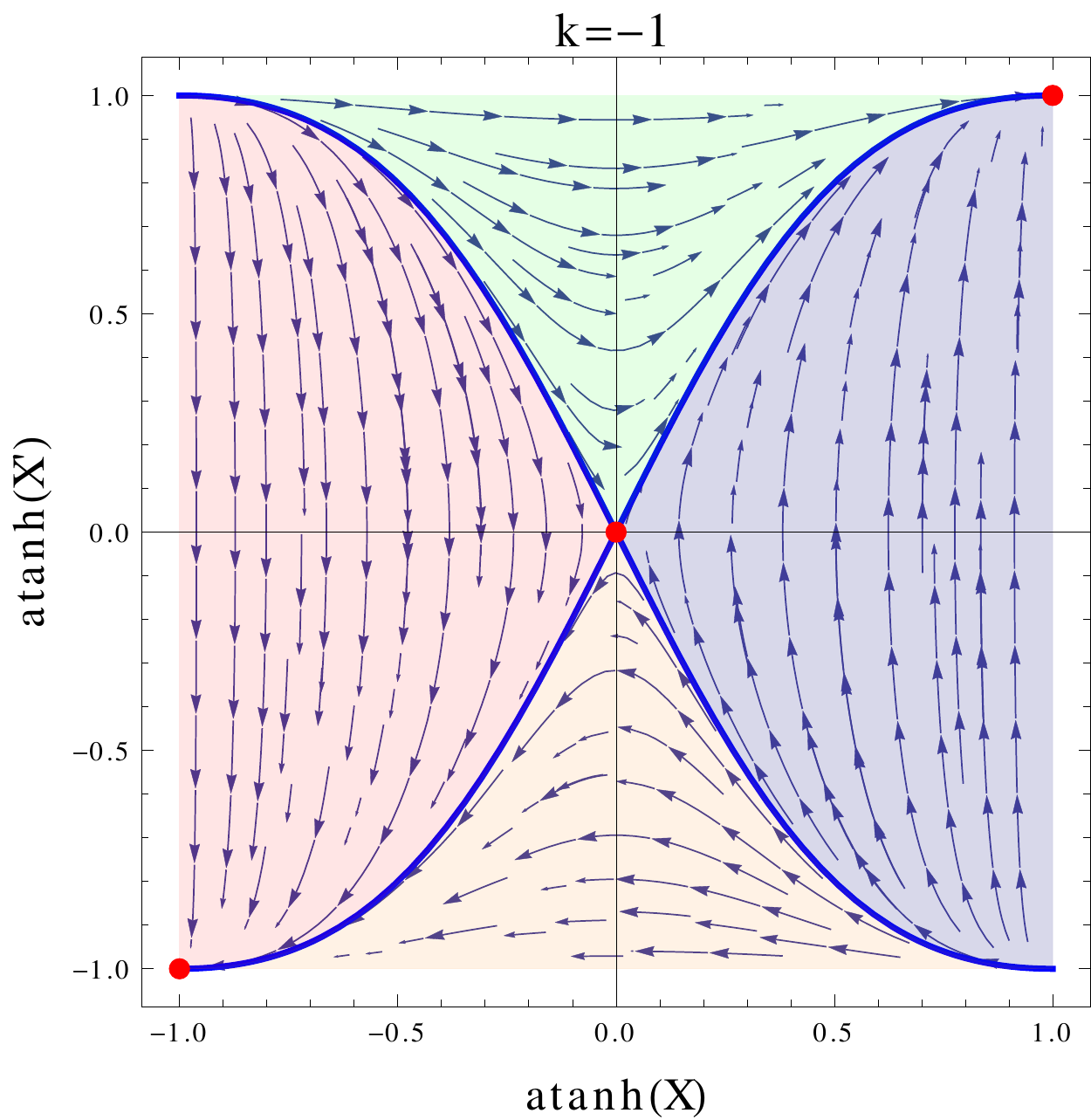}\;\;\;\;\;\;
\includegraphics[height=.20\textheight=.2,angle=0]{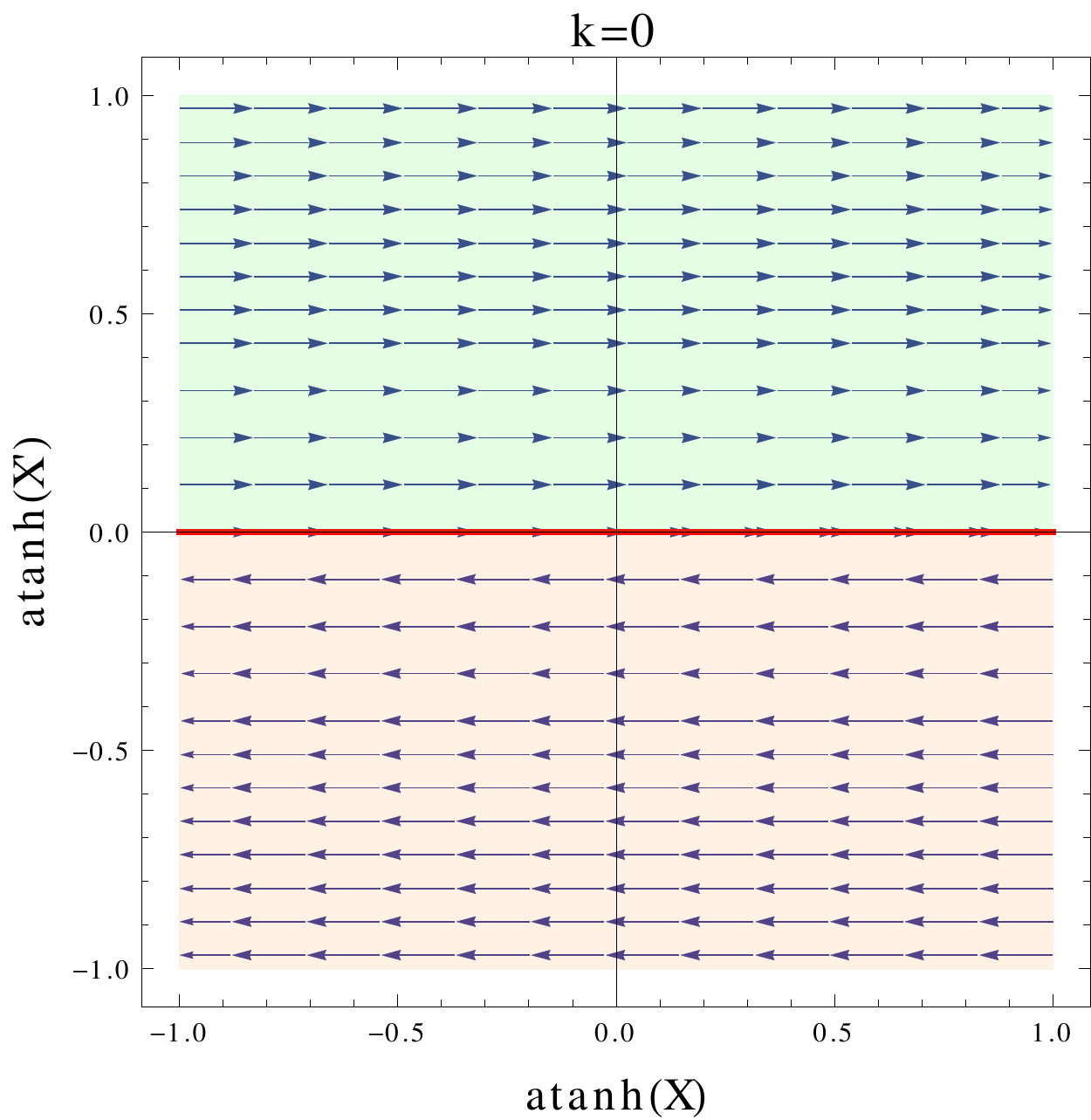}\;\;\;\;\;\;
\includegraphics[height=.20\textheight=.2,angle=0]{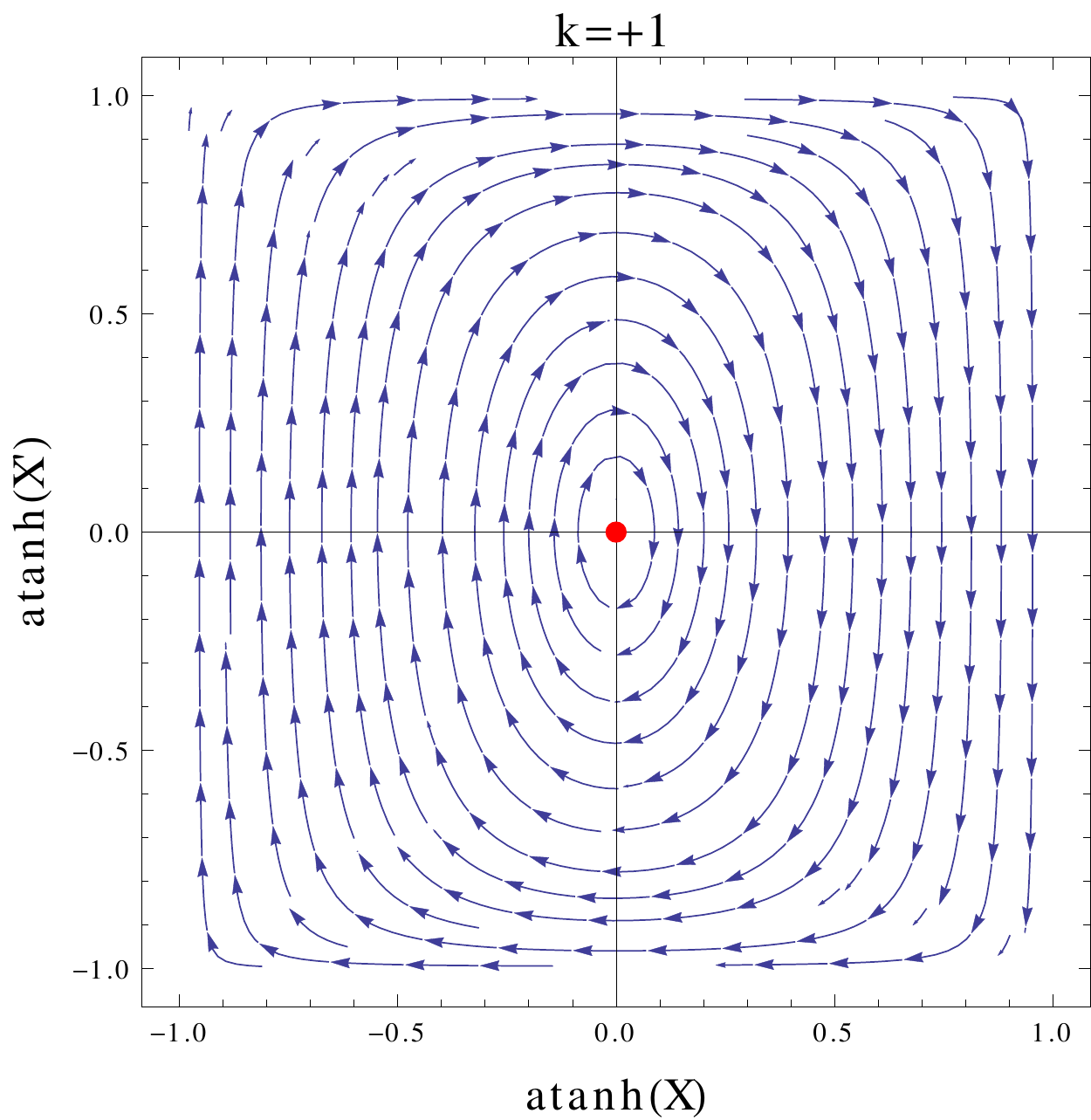}\\
\hspace{+15pt}(\textbf{a})\hspace{+130pt}(\textbf{b})\hspace{+130pt}(\textbf{c})\\ \hspace{+142pt} \\
\includegraphics[height=.20\textheight=.2,angle=0]{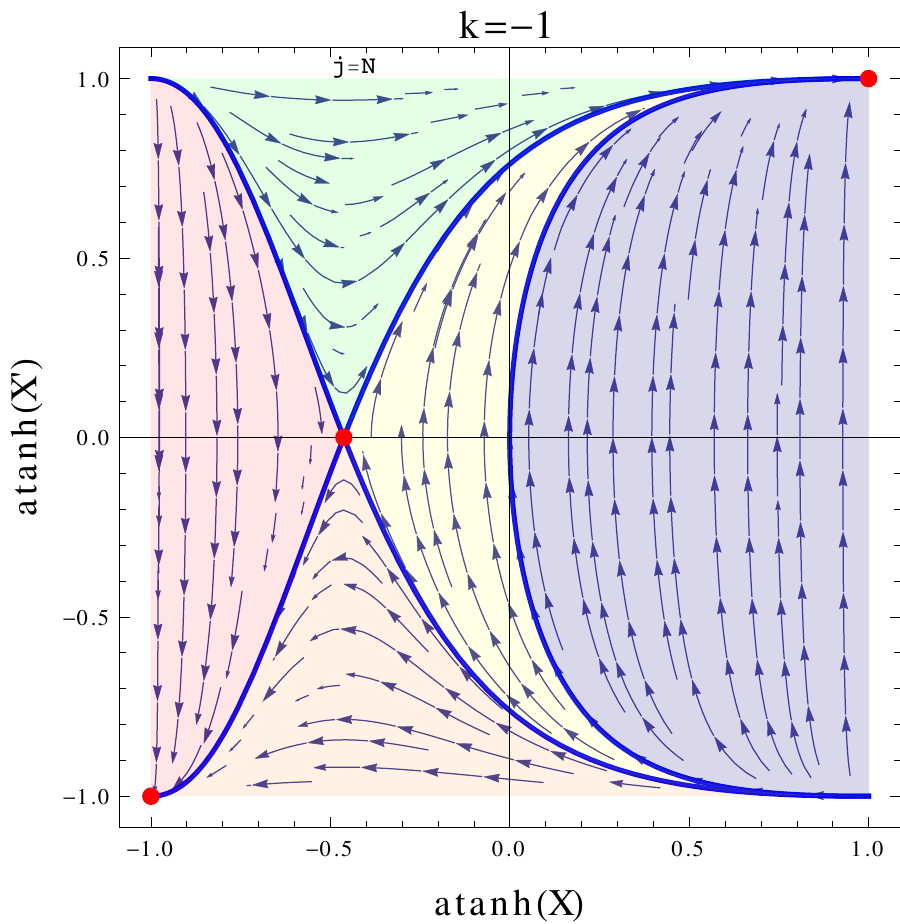}\;\;\;\;\;\;
\includegraphics[height=.20\textheight=.2,angle=0]{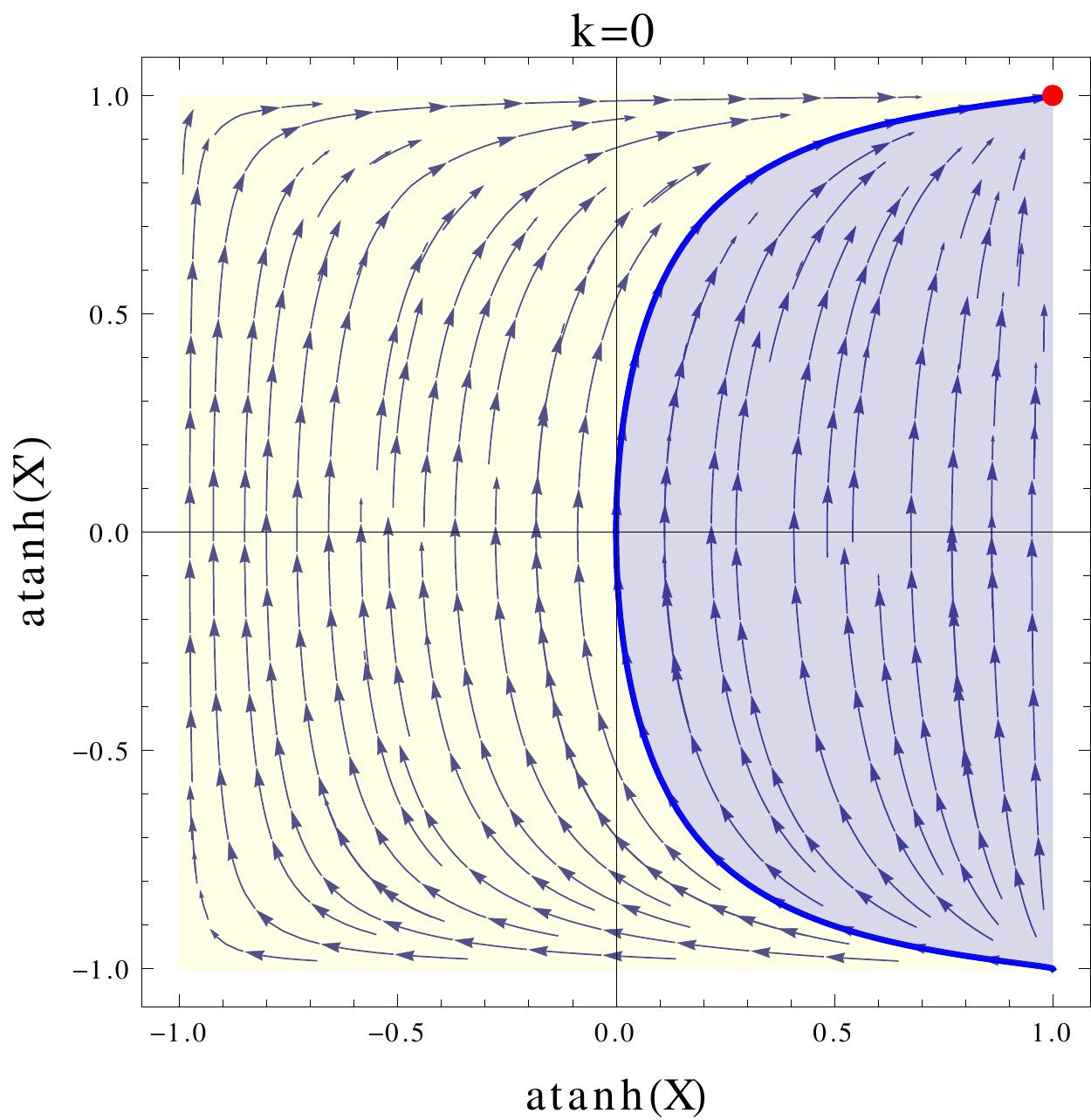}\;\;\;\;\;\;
\includegraphics[height=.20\textheight=.2,angle=0]{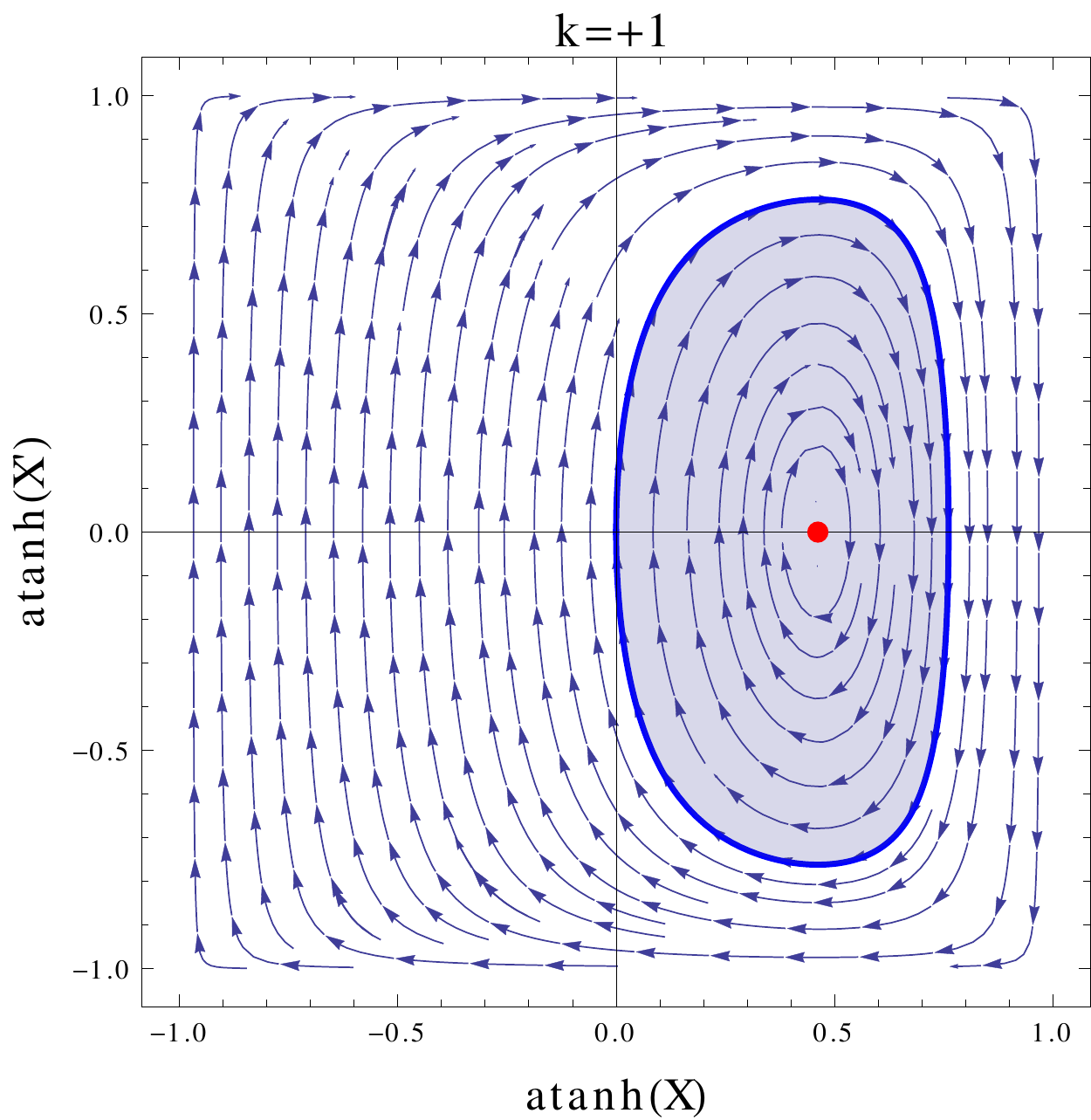}\\
\hspace{+15pt}(\textbf{d})\hspace{+130pt}(\textbf{e})\hspace{+130pt}(\textbf{f})
\caption{Behaviour of $X$ and $X'$ for a scalar-tensor theory without potential. The upper three phase diagrams, (\textbf{a}--\textbf{c}), respectively, correspond to  the $k=-1, 0, +1$ cases for vacuum and stiff fluid. The~lower three phase diagrams, (\textbf{d}--\textbf{f}), respectively, correspond to  the $k=-1, 0, +1$ cases for radiation.}
\label{figureL0}
\end{figure}

The plots represented in Figure \ref{figureL0} show the phase diagrams of this systems $(X, W)$ for the case without potential, i.e., $\lambda_0=0$.  This is again done resorting to  a phase space compactification where  the points at $X, W\to \pm \infty$ are located at the boundaries $x =  \arctanh X = \pm 1$, $y= \arctanh W=\pm 1$. One~realizes that there are trajectories which cross the $X=0$ dividing line in both directions, thus~promoting the transition between a negative $\phi$ into a positive $\phi$, and conversely (and~hence of a swapping of the sign of $G$, as $\phi=G^{-1}$). Please note that as in the GR case previously considered, there is a mirror reflection between the top and lower half of the phase diagrams, arising from time reversal. 

We have used the following color scheme depending on the beginning and the end of the trajectory: (i) Blue: The trajectory starts and finishes with positive value of $\phi$; (ii) Yellow: It begins and finishes with a positive value, but passes through negative values; (iii) Green: It starts with negative values, but finishes with positive values; (iv) White: The trajectory  oscillates between positive and negative values; (v) Orange: It starts with positive values, but finishes with negative value;
(vi) Red: It starts and finishes with  a negative value of $\phi$ (and hence of $G$).

In Figure  \ref{figureL0}, the top three phase diagrams represent vacuum and stiff fluid models, whereas the lower three  correspond to radiation models. From left to right we have $k=-1, 0$  and $k=+1$ models, respectively. It is immediately apparent that in the vacuum models the number of trajectories that cross in one direction, say from left to right,  is the same as that of those which cross in the opposite direction. Once again there is a mirror symmetry with time reversal between the top half and the lower one. Therefore, assuming that a measure of the probability of the model to have a positive gravitational constant, or otherwise, is proportional to the phase space area, we realise that both signs occur with the same probability.  In this sense, the same behaviour occurs for the three cases of vacuum or stiff fluid. In addition, 
in the case of a positive curvature, i.e., Figure \ref{figureL0}c the sign oscillates forever, while in the open models the trajectories evolve towards the Milne solution,  $x=y=1$, 
in the $k = \pm 1$ case, and~are characterized by $X'=\pm f_0$ in the $k = 0$ models, which actually correspond to the solutions found in~\cite{O'Hanlon+Tupper 72}.

On the other hand, for radiation~\cite{GeneralSTsols}, there is a preference for the positive sign, or~at least,  to~finish with a positive gravitational coupling, as represented by the blue and yellow areas. This is a reflection of the fact that the solutions  are late time dominated by the matter component~\cite{Nariai 68,Barrow+Mimoso 94}.  This~is more apparent in the $k=-1,0$ cases, but in the $k=+1$ cases now exhibits  a subset of trajectories with oscillatory behavior, confined in the $\phi>0$ region. The~impact of matter is dependent on the scalar-tensor coupling $\omega(\phi)$. When $1/\sqrt{2\omega+3} \to 0$ the scalar-tensor theories  approach GR, and this is implicit in Equation~(\ref{Y_integrable}).

We thus see that the dynamics of the vacuum FLRW models in ST gravity does not favour  positive values of G with regard to the alternative possibility of a negative G. The~dynamics is such that  the upper half of the phase space corresponds to $G>0$ and the other half to $G<0$, and mirror reflections of one another. Thus both possibilities in what concerns the sign of $G$ have the same probability. However, when matter is present, in the case of open models (including $k = 0$) there is higher probability for a positive value of $G$, following from of a larger proportion of solutions which evolve to become matter dominated.

\subsection{Models with a Cosmological Potential}

\begin{figure}[H]
\centering
\includegraphics[height=.21\textheight=.2,angle=0]{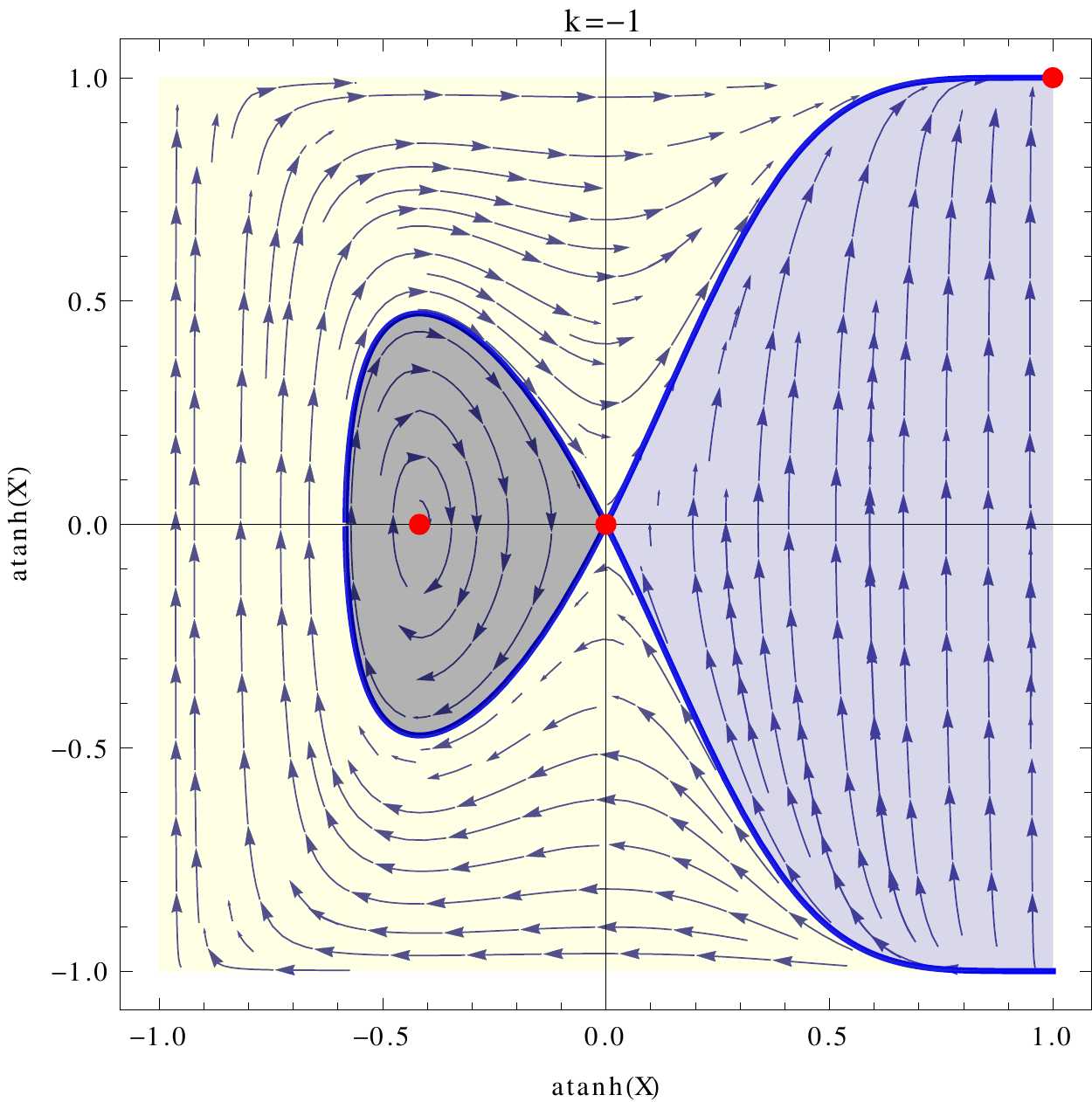}\;
\includegraphics[height=.21\textheight=.2,angle=0]{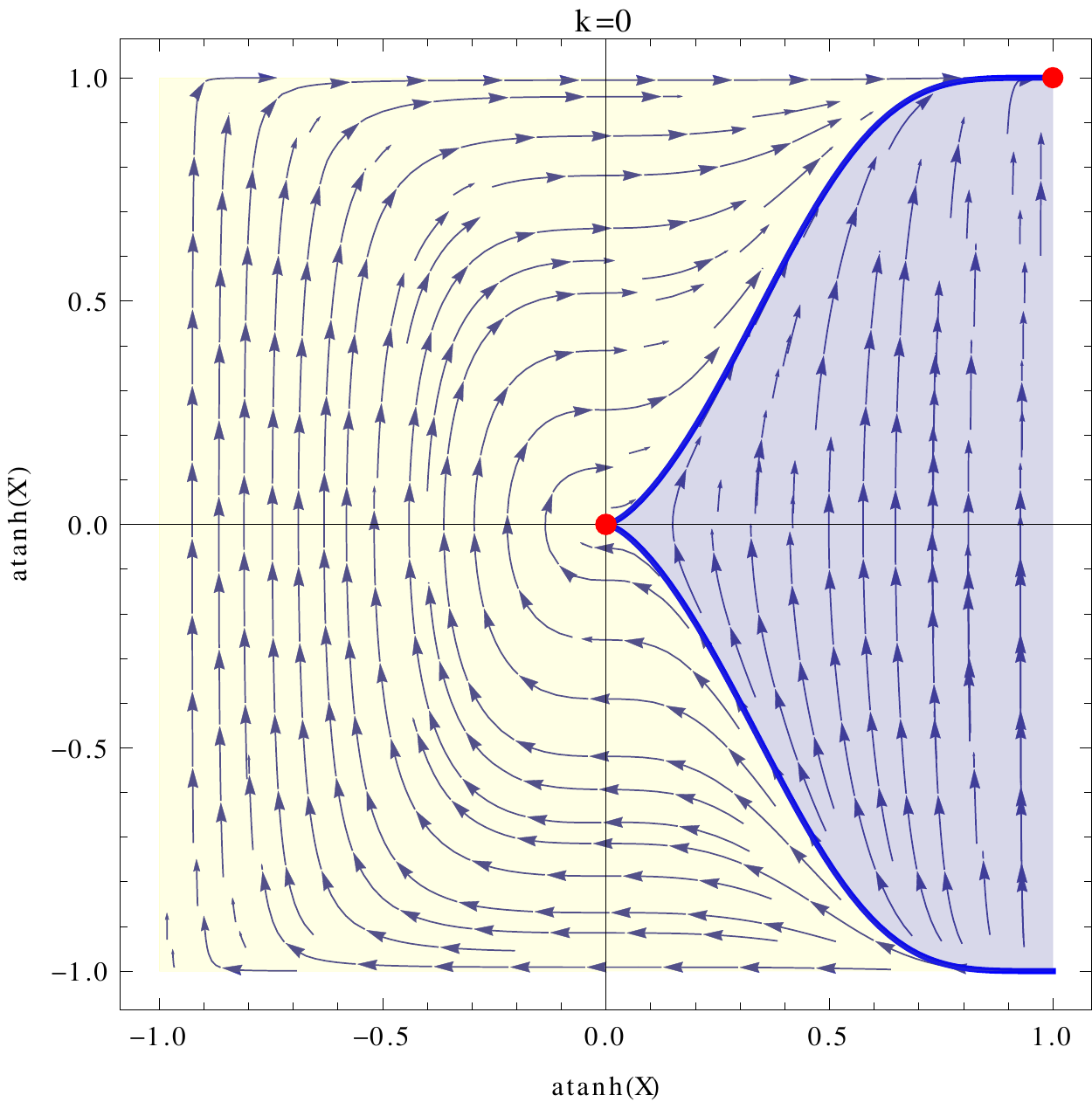}\;
\includegraphics[height=.21\textheight=.2,angle=0]{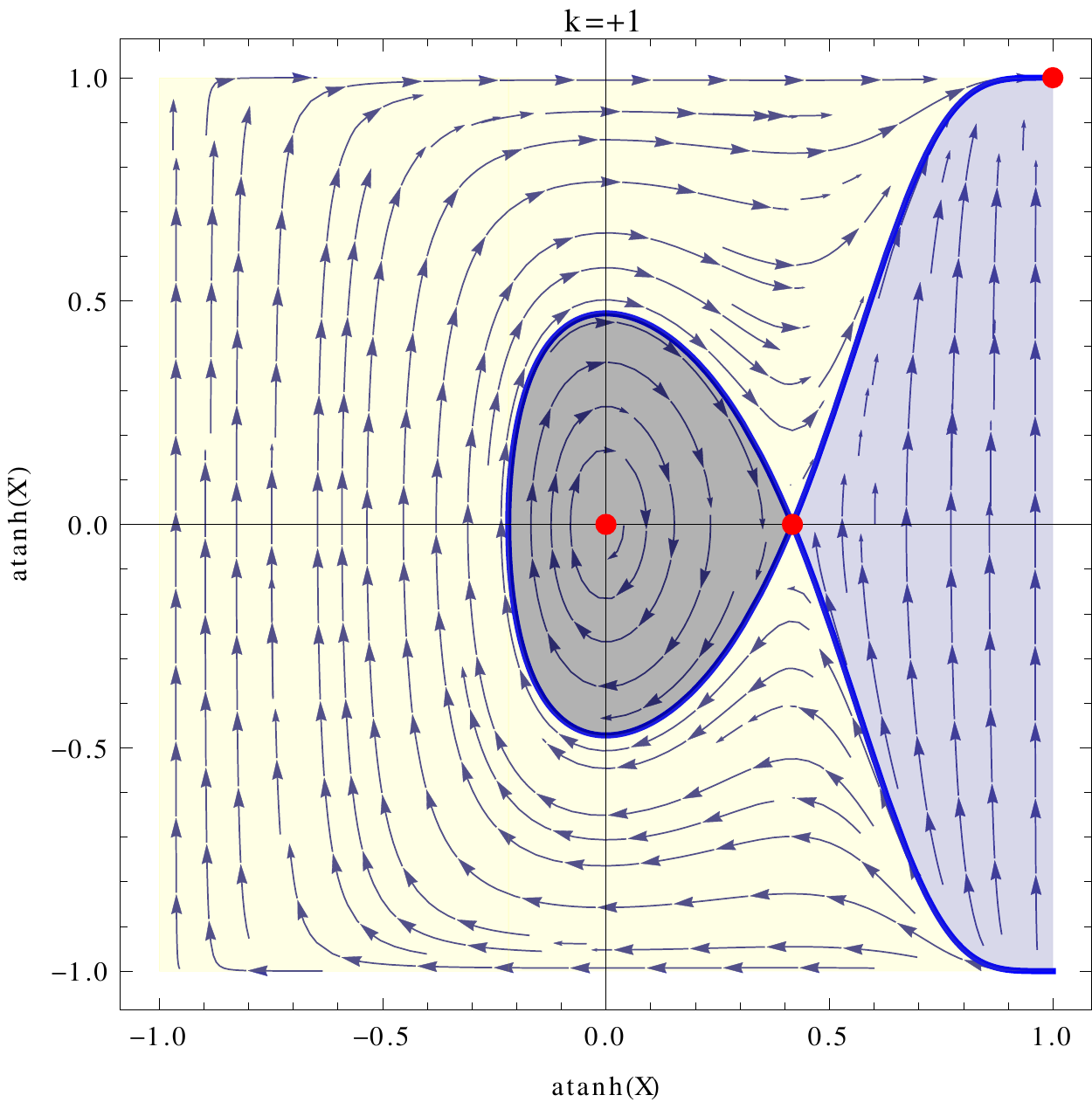}\;\\
\hspace{+13pt}(\textbf{a})\hspace{+123pt}(\textbf{b})\hspace{+123pt}(\textbf{c})\\ \hspace{+142pt} \\
\includegraphics[height=.20\textheight=.2,angle=0]{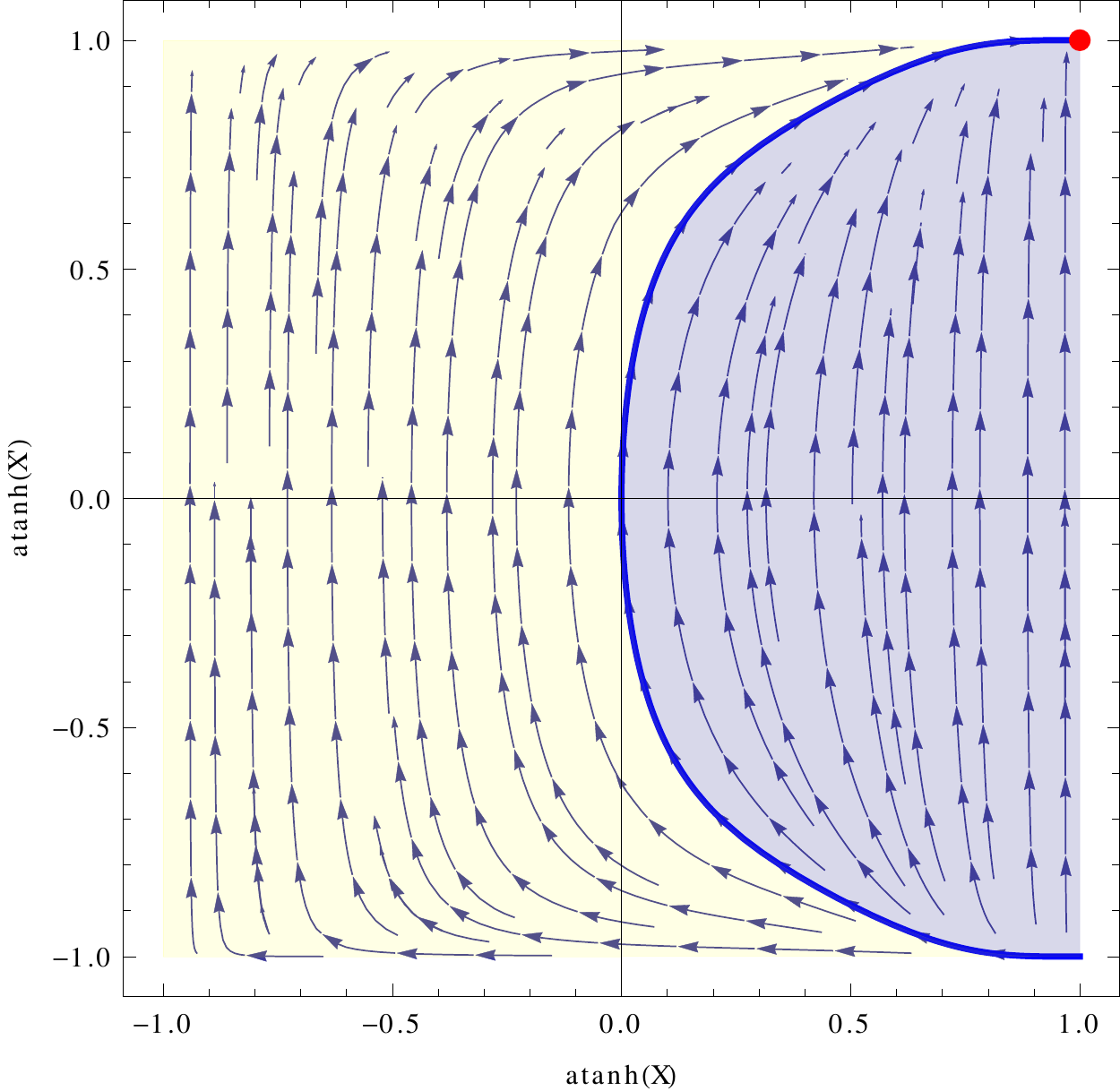}\;
\includegraphics[height=.20\textheight=.2,angle=0]{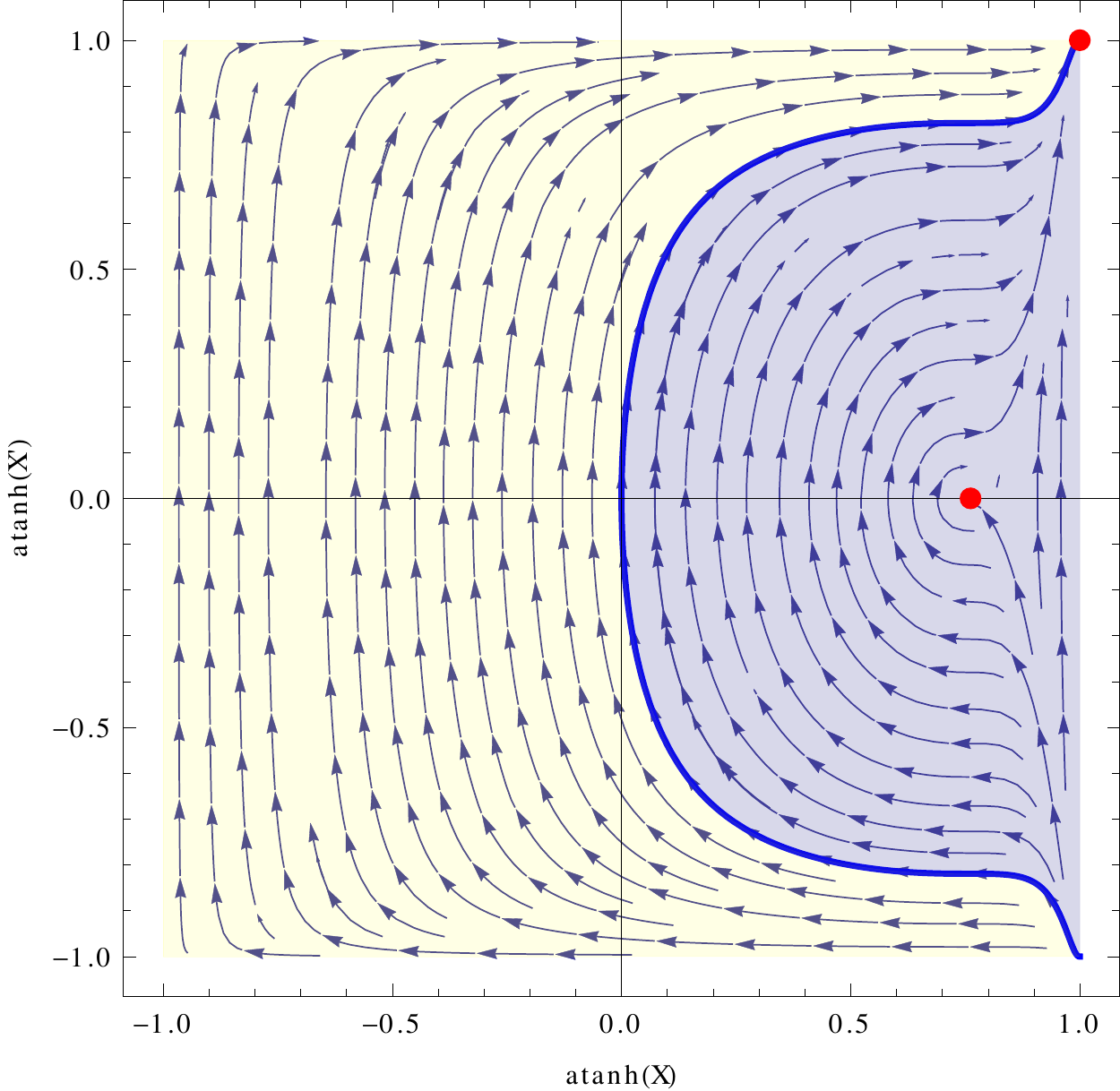}\;
\includegraphics[height=.20\textheight=.2,angle=0]{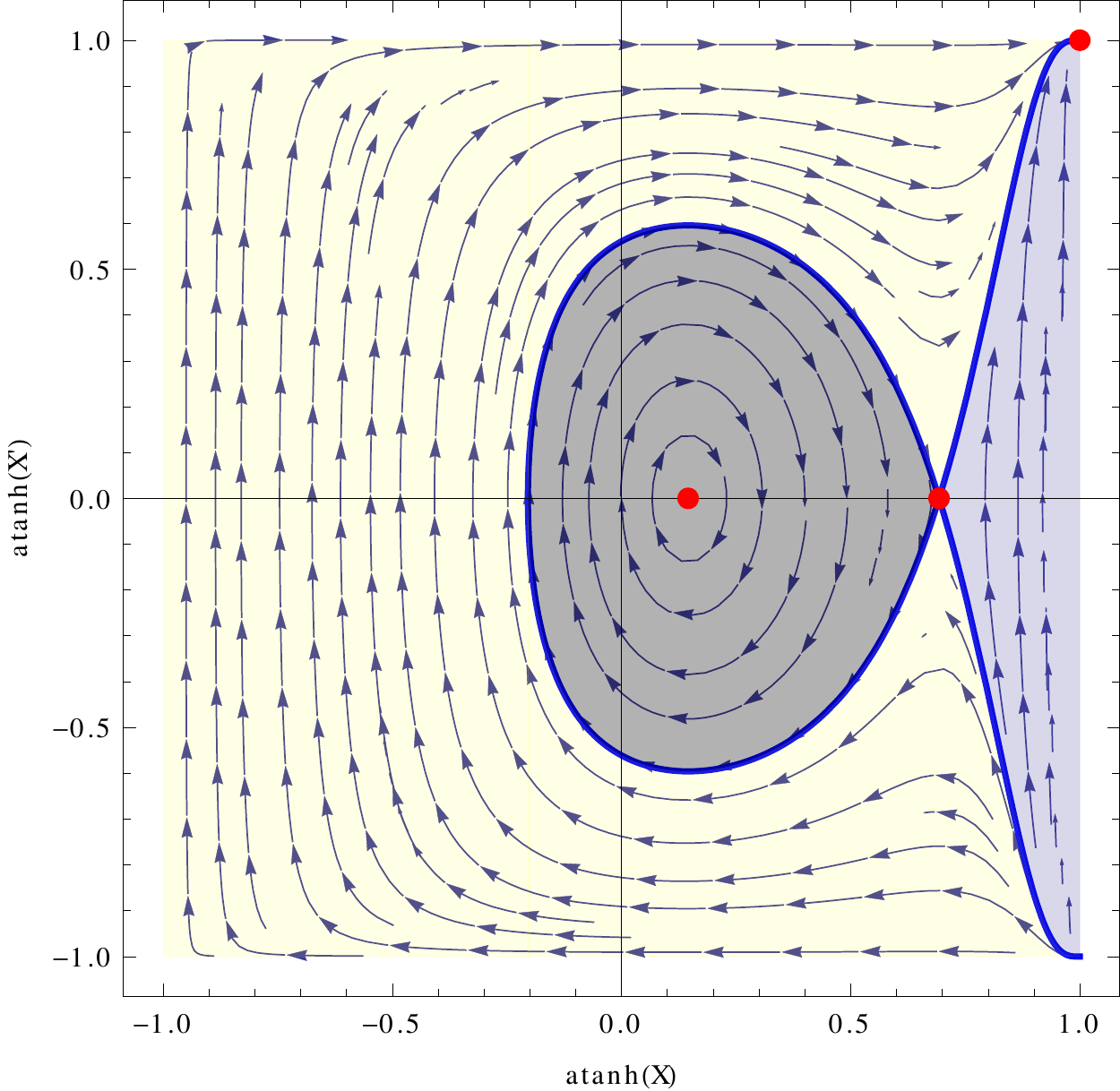}\;\\
\hspace{+13pt}(\textbf{d})\hspace{+123pt}(\textbf{e})\hspace{+123pt}(\textbf{f})
\caption{Behaviour of $X$ and $X'$ for a scalar-tensor theory with a potential $\lambda_0$.
 The upper three phase diagrams, (\textbf{a}--\textbf{c}), respectively, correspond to  the $k=-1, 0, +1$ cases for vacuum and stiff fluid. The~lower three phase diagrams, (\textbf{d}--\textbf{f}), depict the qualitative behaviour of  radiation models, with: (\textbf{d})~When $M\lambda_0>1$ (\textbf{e}) When $M\lambda_0=1$ (\textbf{f}) When $M=1/4$ and $\lambda_0=2$. }
\label{vacuumstiffrad}
\end{figure}

When we allow for a  potential with a positive $\lambda_0$, we see that a quite different picture emerges. Indeed, the phase diagrams corresponding to this case are represented in \ref{vacuumstiffrad}, and we see that now there is an equilibrium point at at $x=1$, $x'=1$, corresponding to $X=+\infty$, $X'=+\infty$, which attracts almost all trajectories of the phase plane, and this happens for all spatial curvature cases. This attractor at infinity  corresponds to a de Sitter attractor, and thus to exponential behaviour of $X$ in cosmic time. 
The only trajectories which do  not end at this critical point are found in the closed $k=+1$ and open models $k=-1$, circling the center equilibrium point.

This is illustrated by the vacuum and radiation which were envisaged in the massless ST models in the previous subsection.

\begin{itemize}
    \item Vacuum $(M=0)$ or stiff fluid $\gamma=2$ with a cosmological potential

 Recalling  that $U(\phi)=\lambda_0\phi^2$, we derive for these two cases:
\begin{eqnarray}
X'&=&W\nonumber\\
W'&=&2\lambda_0 X^2-4 k X
\end{eqnarray}
Please note that the case $\gamma=2$ corresponding to stiff matter can be shown to be reducible to the vacuum case of a theory with a different coupling strength $\omega(\phi)$ (see \cite{GeneralSTsols2}, and the companion paper \cite{GeneralSTsols3} to the present work).

Now, the fixed points $\{X,X'\}$ within a finite locus  will be positioned at  $\{0,0\}$ and $\{2k/\lambda_0,0\}$. To show the graphics of the phase diagrams, $\lambda_0$ has been taken equal to $4,5$ with the purpose to show both points sufficiently separated.

\item Radiation case $\gamma=4/3$ with a cosmological potential

In this case the system is:
\begin{eqnarray}
X'&=&W\nonumber\\
W'&=&2M+2\lambda_0 X^2-4 k X
\end{eqnarray}
and the fixed points are:
\begin{eqnarray}
X= \frac{k\pm \sqrt{k^2-M\lambda_0}}{\lambda_0}, \hspace{1cm} W = 0,\nonumber
\end{eqnarray}

 Therefore, for the cases $k=\pm 1$, at the fixed points we require $M\lambda_0<1$. When this is not satisfied  there are no fixed points, as illustrated in Figure \ref{vacuumstiffrad}d. For the case $k=0$ there are no fixed points within the finite region of the phase plane.
\end{itemize}

The qualitative behaviour depicted in Figure \ref{vacuumstiffrad} reveals that with  the exception of the oscillatory solutions, confined to a closed patch, all other solutions emerge from a collapsing deS solution at $X=\infty, X'=-\infty$ and end in the deS solution at $X=\infty, X'=+\infty$, thus revealing the domination of the cosmological potential term~\cite{Mimoso:1998dn,DynSis_BD1,Capozziello:1994nj,Capozziello:1995kf,Charters:2001hi,Hrycyna:2013yia,Garcia-Salcedo:2015naa}.  More importantly, we see that the solutions are attracted to the positive $G$  half plane.  We thus conclude that the consideration of a cosmological potential\footnote{One possible origin for such a potential might be found from a mechanism similar to the dark fluid model of \cite{Luongo:2018lgy}.} has the power to induce  the dynamics of FLRW models to favour  positive values of $G$ instead of negative $Gs$. Thus this provides a cosmological mechanism to stabilize $G$ in the positive sector. In addition, it happens  that the deS asymptotic behaviour is accompanied with a relaxation towards GR~\cite{Mimoso:1998dn,Mimoso:1999ai,Mimoso:2003iha,Jarv:2010zc,DynSis_GenSTT}

\section{Observational Features}

ST gravity theories have to satisfy  several observational  bounds, namely, the so-called Parametrized Post-Newtonian (PPN) weak field, solar system tests, bounds stemming from the cosmic microwave background (CMB), from baryonic acoustic oscillations (BAO), and from primordial Big-Bang Nucleosynthesis (BBN), as well as  bounds on the time variation of the  \mbox{gravitational ``constant'' $\dot G/G$ \cite{Will:1986nk,Ooba:2017gyn,Will:2014kxa,Koyama:2015vza}. }

The present state of the art tell us that at present $\omega_P \gtrsim 4\times 10^4$, $\dot G/G\lesssim 10^{-10} \,{\rm yr}^{-1}$, and from BBN we have  $\alpha_0^2\lesssim 10^{-6.5}\,\beta^{-a}\,\left(\Omega_{matter}h_0 \right)^{-3/2} $  when $\beta>0.5$~\cite{Mimoso:1992vr,
Damour:1998ae,Larena:2005tu,Iocco:2008va,Coc:2012xk}, where $\alpha$ and $\beta$ correspond to the  Damour and Nordtvedt (DN)~\cite{Damour:1993id} parametrization $(\sqrt{2\omega+3})^{-1}= \alpha(\tilde\phi)$, with $\alpha(\tilde\phi)= \alpha_0+\beta_0\,(\tilde\phi-\tilde\phi_0)$, $\tilde\phi(\phi) =\int\sqrt{(2\omega+3)/2}\; {\rm d}\ln\phi$, being the redefined BD scalar field in the so-called Einstein frame. Indeed, as shown by DN,  it is possible to associate a second internal potential ${A}(\tilde \phi)$ to the running coupling $\omega(\phi)$ such $\alpha(\tilde\phi)=d{A}/d\tilde\phi$ and whenever it has a minimum it drives the relaxation to GR.

The local weak field bounds can be somewhat alleviated if some chameleon or Vainshtein mechanism  applies, but it is  difficult to evade the other bounds on  wider scales. 
Yet, the vast majority of these bounds pertain to models where the scalar field has no cosmological potential (see though~\cite{Coc:2012xk}). 


These bounds, therefore, imply that a primordial variation of the gravitational coupling must have been severely damped before the time of BBN such that a positive coupling, satisfying mild deviations from GR \cite{Larena:2005tu} is not only compatible with the observations on light elements abundances, but also solves the so-called $^7$Li problem. In the suite, during the following radiation and matter epochs, the cosmological approach to GR is achieved, implying that $G$ is positive. 

 The mechanism investigated in this work fulfills this scenario. It tells us that the solutions approach a de Sitter behaviour, which corresponds to $X\propto (\eta_0-\eta)^{-2}$ and $X'\propto (\eta_0-\eta)^{-3}$, so that $X(t)\propto \exp( \sqrt{\lambda_0/3}\; \tilde t)$. When we consider the radiation epoch of the universe, we have 
 \begin{equation}
Y'=\frac{f_0}{X(\eta)},
 \end{equation}  
 and hence, $Y'\to 0$, asymptotically with $X\to + \infty$. As $Y'=\sqrt{(2\omega(\phi)+3)/3}\;\phi'/\phi$, we obtain in cosmic time (in the so-called Einstein frame)
  \begin{equation}
\frac{\dot G}{G} = \pm \sqrt{\frac{3}{2\omega+3}} \,\frac{f_0}{\sqrt{X}} \equiv \pm \,\frac{\sqrt{3}f_0\,\alpha(\phi)}{\sqrt{X}}\,.
 \end{equation}
 
As $Y' = f_0/X \to 0$, and  the PPN parameter $\alpha\propto (\sqrt{2\omega(\phi)+3})^{-1}$  satisfies  the bound $\alpha\lesssim 10^{-8}$, as~$\omega_P \gtrsim 4\times 10^4$, we have that $\dot G/G\lesssim 10^{-8}/\sqrt{X(\tilde t)}$, where $X(\tilde t) \to \infty$ in an exponential way, thus easily meeting the bounds on $\dot G/G$. Please note that in this limit, $G\to G_N$ and $X(t)$ also becomes exponential in the Jordan frame.  
%



 

\section{Summary and Conclusions}

In this work, we have investigated a cosmological mechanism that induces the value of the gravitational effective coupling ``constant'' to be positive. This is naturally done in the framework of scalar-tensor (ST) gravity theories, where this coupling varies, and which thus allow for
the possibility of a negative coupling. We have considered the cosmological evolution of ST models both with and without the presence of a cosmological potential. We have resorted to a dynamical systems analysis which enable us to put in evidence the relevant qualitative features of the models. In the absence of the cosmological potential, the presence of matter or radiation favours a positive value of the gravitational ``constant'', when the evolution enters a phase of matter domination. This a mild effect and it is a consequence of Damour and Nordtvedt's  relaxation mechanism towards GR~\cite{Damour:1993id}.  However, it is when a quadratic cosmological potential, $U(\phi)=\lambda_0\,\phi^2$, is present that an  attracting mechanism towards a positive value of the gravitational running ``constant'' becomes manifest. This is accompanied by  an asymptotic de Sitter behaviour. 

 By the same token, this system produces two additional effects: a de Sitter inflation and a relaxation towards general relativity. The~latter effect allows, in particular, the fulfilment of the observational bounds on $|\dot G/G|$,
when the potential is exactly quadratic in the Jordan frame. It~effectively acts as a cosmological constant in the Einstein frame and the stabilization of the gravitational constant in the positive sector, may be seen as a by-product  of the cosmic no-hair theorem. This mechanism of stabilization of the sign of $G$ should take place early enough, in the primordial stages of the universe, consistently with the latest  assessments of observational constraints on ST theories~\cite{Lee:2010zy,Ooba:2017gyn}

\vspace{6pt}


\authorcontributions{These authors contributed equally to this work.}

\funding{This research was funded by Funda\c{c}\~ao para a Ci\^encia e a Tecnologia (FCT) grant number PD/BD/114435/2016 under the IDPASC PhD Program and also grant number UID/FIS/04434/2013. This research was also funded by the projects PTDC/FIS- OUT/29048/2017 and IF/00852/2015. of Funda\c c\~ao para a Ci\^encia e a Tecnologia \url{https://www.fct.pt/index.phtml.en.}

}

\acknowledgments{JPM expresses his gratitude to Norbert Van den Bergh for kindly bringing the references~\mbox{\cite{Roxburgh:1980,Roxburgh:1980b}} to his attention. In the suite all of the authors  are also  indebted to both Prof. Ian Roxburgh and George Rideout of the   Gravity Research Foundation for most kindly enabling the access to these manuscripts.

}

\conflictsofinterest{The authors declare no conflict of interest.} 
\newpage
\abbreviations{The following abbreviations are used in this manuscript:\\

\noindent 
\begin{tabular}{@{}ll}
MDPI & Multidisciplinary Digital Publishing Institute\\
DOAJ & Directory of open access journals\\
GR & General Relativity\\
ST & Scalar-Tensor\\
FLRW & Friedmann-Lemaître-Robertson-Walker\\
BD & Brans-Dicke\\
deS & de Sitter\\
DN & Damour and Nordtvedt\\
PPN & Parametrised Post-Newtonian \\
CMB & Cosmic microwave background\\
BBN & Big-Bang nucleosynthesis
\end{tabular}}


\reftitle{References}





\end{document}